\documentclass[prd,amsfonts,nofootinbib,showpacs]{revtex4}
\pdfoutput=1
\usepackage{setspace}
\usepackage{epsfig}
\usepackage{psfrag}
\usepackage{enumitem}
\usepackage{latexsym}
\usepackage{indentfirst}
\usepackage{fancyhdr}
\usepackage{amssymb}
\usepackage{amsmath}
\usepackage{amsfonts}
\usepackage{pifont}
\usepackage{bbold}

\usepackage{color}

\usepackage{graphicx}
\usepackage[center,footnotesize,hang]{subfigure}
\usepackage{url}
\usepackage{array}
\usepackage{natbib}
\setcitestyle{aysep={}}

\DeclareMathAlphabet{\mathsc}{OT1}{cmr}{m}{sc}
\newcommand {\ignore}[1]{}
\renewcommand{\baselinestretch}{1}

\newcommand{\nn}  {\nonumber}

\def\10{$SO(10)$}
\def\21{SU(2) $\otimes$ U(1) }

\def\422{$SU(4) \otimes SU(2) \otimes SU(2)$}
\def\321{SU(3) $\otimes$ SU(2) $\otimes$ U(1)}
\def\gsim{\raise0.3ex\hbox{$\;>$\kern-0.75em\raise-1.1ex\hbox{$\sim\;$}}}
\def\lsim{\raise0.3ex\hbox{$\;<$\kern-0.75em\raise-1.1ex\hbox{$\sim\;$}}}

\def\lsim{\raise0.3ex\hbox{$\;<$\kern-0.75em\raise-1.1ex\hbox{$\sim\;$}}}
\def\gsim{\raise0.3ex\hbox{$\;>$\kern-0.75em\raise-1.1ex\hbox{$\sim\;$}}}
\def\vev#1{\left\langle #1\right\rangle}
\def \znbb {0\nu\beta\beta}
\renewcommand{\baselinestretch}{1.25}
\newcommand{\AddrAHEP}{%
  AHEP Group, Institut de F\'{\i}sica Corpuscular --
  C.S.I.C./Universitat de Val{\`e}ncia \\
  Edificio Institutos de Paterna, Apt 22085, E--46071 Valencia, Spain}

\newcommand{\ba}{\begin{array}}
\newcommand{\ea}{\end{array}}
\relax
\def\321{$SU(3)\times SU(2)\times U(1)$}
\begin{document}



\renewcommand{\Huge}{\Large} \renewcommand{\LARGE}{\Large}
\renewcommand{\Large}{\large} 
\def \znbb {$0\nu\beta\beta$ }

\title{Constraining neutrinoless double beta decay}


\author{L.~Dorame}\email{dorame@ific.uv.es }\affiliation{$^{1}$\AddrAHEP}  
\author{D.~Meloni}\email{meloni@fis.uniroma3.it}\affiliation{$^{2}$ Dipartimento di Fisica "E. Amaldi", \\
Universit\'a degli Studi Roma Tre, Via della Vasca Navale 84, 00146 Roma, Italy}  
\author{S.~Morisi}\email{morisi@ific.uv.es}\affiliation{$^{1}$\AddrAHEP}
\author{E.~Peinado}\email{epeinado@ific.uv.es}\affiliation{$^{1}$\AddrAHEP}
\author{J.~W.~F.~Valle} \email{valle@ific.uv.es}\affiliation{$^{1}$\AddrAHEP}

\date{\today}
\renewcommand{\baselinestretch}{1.8}
\begin{abstract}

  A class of discrete flavor-symmetry-based models predicts
  constrained neutrino mass matrix schemes that lead to specific
  neutrino mass sum-rules (MSR).  We show how these theories may
  constrain the absolute scale of neutrino mass, leading in most of
  the cases to a lower bound on the neutrinoless double beta decay
  effective amplitude.
\end{abstract}
\maketitle

\section{Introduction}
\label{intro}

The discovery of
oscillations~\cite{nakamura2010review,Abe:2011sj,Double-Chooz,Schwetz:2011qt,Schwetz:2011zk,Fogli:2011qn,Maltoni11,schwetz:2008er}
implies non-vanishing neutrino masses and mixing providing one of the
most solid indications for physics beyond the Standard Model. The
fact that neutrinos have very tiny masses, in contrast to charged
leptons and quarks, and that two of the mixing angles are large, are
among the deepest theoretical puzzles in particle physics. Since
neutrinos carry no electric charge, they are expected on general
grounds to be Majorana particles~\cite{schechter:1980gr}, leading to
the existence of lepton number violating
processes~\cite{schechter:1982bd,Duerr:2011zd}.
This intriguing possibility will be hopefully confirmed by the
observation of neutrinoless double beta decay ($0\nu\beta\beta$)
processes~\cite{Barabash:2011fn,Rodejohann:2011mu}.  Indeed, upcoming
$0\nu\beta\beta$ experiments are expected to improve the sensitivity
by up to about one order of magnitude~\cite{Angrik:2005ep,:2009yi,Abt:2004yk,Alessandrello:2002sj}.

It seems unlikely that the observed pattern of lepton mixing angles is
an accident: it probably indicates the existence of an underlying
flavor symmetry of some sort, either an Abelian
symmetry~\cite{Froggatt:1978nt} or a non-Abelian
one~\cite{Ishimori:2010au}. In the former case one typically obtains
texture zeros for the mass
matrices~\cite{fritzsch:1979zq,fritzsch:2002ga,hirsch:2007kh} but is
unable to predict mixing angles.
In contrast, non-Abelian flavor symmetries are potentially more
powerful, allowing also in principle for mixing angle predictions.
As an example, several realizations of non-Abelian discrete flavor
symmetry schemes lead to an effective neutrino mass matrix which
corresponds to a numerical (parameter-free) prediction for lepton
mixing. A popular example of such neutrino mass matrix $M^\nu$ is the
tri-bimaximal (TBM)~\cite{Harrison:2002er} type~\footnote{There are
  also non-Abelian discrete flavor symmetry schemes leading to the
  bi-maximal lepton mixing
  pattern~\cite{Mohapatra:1998ka,Barger:1998ta,Grimus:2008tt,Altarelli:2009gn,Meloni:2011fx},
  as well as golden-ratio
  schemes~\cite{Datta:2003qg,Kajiyama:2007gx,Everett:2008et,Ding:2011cm}.
}, characterized by
\begin{equation}\label{Mtbm}
M^\nu=M_{TBM}\equiv \left(
\begin{array}{ccc}
x & y & y\\
y&x +z& y-z\\
y& y-z& x+z
\end{array}\right),
\end{equation}
which depends only on three complex parameters $x$, $y$ and $z$.
In the mass eigenstate basis the three complex parameters $x$, $y$
and $z$ would correspond to three neutrino mass parameters plus two
Majorana CP phases~\cite{schechter:1980gr,Rodejohann:2011vc}, as the
Dirac phase disappears since $\theta_{13}=0$.
Many such schemes are characterized by a specific (complex) relation
among the parameters $x$, $y$ and
$z$~\cite{Ma:2005sha,Altarelli:2005yp,Altarelli:2005yx,Altarelli:2006kg,Ma:2006vq,Bazzocchi:2007na,Bazzocchi:2007au,Honda:2008rs,
  Brahmachari:2008fn,Lin:2008aj,Chen:2009um,Ma:2009wi,Fukuyama:2010mz,Bazzocchi:2008ej,Chen:2007afa,Ding:2008rj,Chen:2009gy,Barry:2010zk,Ding:2010pc,Bazzocchi:2009da,
  Bazzocchi:2009pv,Burrows:2010wz,Babu:2005se,He:2006dk,Morisi:2007ft,Altarelli:2008bg,Adhikary:2008au,Csaki:2008qq,Altarelli:2009kr,Lin:2009bw,Hagedorn:2009jy,
  Burrows:2009pi,Berger:2009tt,Ding:2009gh,Mitra:2009jj,delAguila:2010vg,Ding:2011gt},
leaving only two free complex parameters, further reducing the number
of independent model parameters describing the lepton sector.
In the mass basis these correspond to only two independent neutrino
mass eigenvalues (the other follows from the existence of a neutrino
mass sum rule), plus two Majorana CP phases (as mentioned, the Dirac
phase is unphysical).

In this paper we study the implications of these sum-rule schemes for
the lower bound on the parameter $|m_{ee}|$ characterizing the
amplitude for neutrinoless double beta decay.  We show that, given
that two neutrino mass squared splittings are well-determined by
neutrino oscillation data~\cite{Schwetz:2011qt,Schwetz:2011zk}, we are
left, approximately, with a one-parameter family of neutrinoless
double beta decay theories in which the corresponding amplitude is
mainly determined just by the overall absolute neutrino mass scale.
The paper is organized as follows: in Section II we present the mass
relations; in section III, we obtain the lower limit on $|m_{ee}|$ for
all models considered here and briefly discuss their phenomenological
implications, whereas in Section IV we present our conclusions.

\section{Mass relations}

In this section we focus on a general sub-class of mass matrices
leading to a numerical (parameter-free) prediction for the lepton
mixing matrix, consistent with current neutrino oscillation
data~\cite{Schwetz:2011qt,Schwetz:2011zk} where the following types of
mass relations hold:
\begin{eqnarray}
&A)&\chi\, m_2^\nu+\xi\, m_3^\nu=m_1^\nu, \label{i}\\
&B)&\frac{\chi }{m_2^\nu}+ \frac{\xi}{m_3^\nu}=\frac{1}{m_1^\nu},\label{ii}\\
&C)&\chi\,\sqrt{ m_2^\nu}+\xi\, \sqrt{m_3^\nu}=\sqrt{m_1^\nu}~. \label{iii}
\end{eqnarray}
Here $m_i^\nu=m_i^0$ denote neutrino mass eigenvalues, up to a
Majorana phase factor, while $\chi$ and $\xi$ are free parameters
which specify the model, taken to be positive without loss of
generality. For the sake of completeness, we also consider a fourth
mass relation,
\begin{eqnarray}
&D)&\frac{\chi}{\sqrt{ m_2^\nu}}+\frac{\xi }{\sqrt{m_3^\nu}}=\frac{1}{\sqrt{m_1^\nu}}\,. \label{iiii}
\end{eqnarray}
As far as we can tell, this last relation has not yet been considered
in the literature.
In the following we show how this class of mass matrices arises in
non-Abelian discrete flavor symmetry schemes.  We first consider an
effective dimension-five operator description~\cite{weinberg:1980bf},
and then discuss the cases where the neutrino mass matrix $M^\nu$
arises from various seesaw mechanism realizations, such as type-I or
type-II
seesaw~\cite{schechter:1980gr,minkowski:1977sc,gell-mann:1980vs,yanagida:1979,mohapatra:1981yp,schechter:1982cv,lazarides:1980nt}
or, from alternative low-scale seesaw schemes, for example, the
inverse seesaw
~\cite{Mohapatra:1986bd,GonzalezGarcia:1988rw,akhmedov:1995ip,Akhmedov:1995vm}.

\subsection*{Effective dimension-five operator description}
\noindent

Consider first the dimension five operators $(L L H H) $, where the
parentheses indicate all possible contractions among the irreducible
representations of the underlying unspecified non-Abelian flavor
symmetry group $L$ and $H$ belong to.  Since we want a mass matrix
with only two independent parameters, we assume that our effective
Lagrangian contains only two terms, associated to two independent
field contractions:
\begin{equation}
\mathcal{L}=\frac{y_a}{M} \epsilon_{ij}^a (L_i L_j)_a H H +\frac{y_b}{M} \epsilon_{ij}^b (L_i L_j)_b H H;
\end{equation}
here $M$ denotes the effective scale, $a,b$ represent the two
contractions, while $\epsilon_{ij}^a$ and $\epsilon_{ij}^b$ are
Clebsch-Gordan (CG) coefficients involving relevant field components
$L_{i,j}$.
After electroweak symmetry breaking, the effective neutrino mass
matrix elements are given as linear combinations involving only the
two parameters $a$ and $b$,
\begin{equation}
\label{mma}
M^\nu_{ij}=a\, \epsilon_{ij}^a  +b\, \epsilon_{ij}^b ~,
\end{equation}
where $a=y_a \vev{H}^2/M$ and $b=y_b \vev{H}^2/M$. 

A number of non-Abelian discrete flavor symmetry realizations lead to
the TBM structure for the effective neutrino mass matrix $M^\nu$ in
Eq.~(\ref{Mtbm}) with some suitable relation among the $x$, $y$ and
$z$ coefficients.
It is clear that in this case the corresponding mass eigenvalues will
always be expressed as linear a combination of $a$ and $b$ and hence
will be related to each other through a relation of type $(A)$ in
Eq.~(\ref{i}).

\subsection*{Type-I seesaw mechanism}
\noindent

For the type-I seesaw mechanism there are two simple ways to get a
neutrino mass matrix similar to $M_{TBM}$ depending only upon two free
complex parameters.  In the first case, the Dirac neutrino mass matrix
$m_D$ has the structure given in eq.~(\ref{Mtbm}) while the
right-handed neutrino mass matrix $M_R$ is proportional to a numerical
$\mu-\tau$ invariant matrix satisfying the relation
$(2,2)+(2,3)=(1,1)+(1,2)$ among its elements, like for instance in
Eq.~(\ref{tbmtype}):
\begin{equation}~\label{tbmtype}
M_R\propto \left(
\begin{array}{ccc}
1 & 0 & 0\\
0 & 1 & 0\\
0 & 0 & 1
\end{array}
\right),\quad
\left(
\begin{array}{ccc}
1 & -1 & -1\\
-1 & 1 & -1\\
-1 & -1 & 1
\end{array}
\right),\quad
\left(
\begin{array}{ccc}
2 & 1 & 1\\
1 & 5 & -2\\
1 & -2 & 5
\end{array}
\right),...
\end{equation}
In what follows we call such matrix generically as ``TBM-type''.
It is not difficult to verify that the light neutrino mass matrix
arising from the type-I seesaw formula~\cite{schechter:1981cv}
$M^\nu=-m_D M_R^{-1} m_D^T$ has mass eigenvalues of type $m_i^\nu
\propto (\alpha_i a+\beta_i b)^2$, yielding mass relations of type
{\it (C)}.  For instance, in Ref.\,\cite{Hirsch:2008rp} it has been
found that $m_1^\nu\propto (a+b)^2$, $m_2^\nu\propto a^2$ and
$m_3^\nu\propto (a-b)^2$, from which the relation
$\sqrt{m_1^\nu}+\sqrt{m_3^\nu}=2 \sqrt{m_2^\nu}$ holds.

The second possibility arises when $M_R\sim M_{TBM}$ as in
Eq.~(\ref{Mtbm}), while the Dirac neutrino mass matrix is a numerical
``TBM-type'' matrix, as in eq.~(\ref{tbmtype}). In this case it is
simple to show that the eigenvalues of $M^\nu$ are of the form
$m_i^\nu \propto 1/(\alpha_i a+\beta_i b)$, where $\alpha_i$ and
$\beta_i$ are numerical coefficients, giving a mass relation of type
{\it (B)}.  For instance, in the model of
Ref.\,\cite{Altarelli:2005yp} the authors found $m_1^\nu\propto
1/(a+b)$, $m_2^\nu\propto 1/a$ and $m_3^\nu\propto 1/(b-a)$ from which
the relation $1/m_1^\nu-1/m_3^\nu=2/m_2^\nu$ is satisfied.

\subsection*{Other seesaw mechanisms}
\noindent

Similar conclusions can be obtained for different seesaw mechanisms,
such as type-II.  From the point of view of our classification,
type-II seesaw is equivalent to the dimension five operator case {\it
  (A)}.  

We now move to the inverse seesaw
mechanism~\cite{Mohapatra:1986bd,GonzalezGarcia:1988rw}, which arises
when introducing a fermion singlet $S$ with opposite lepton number
with respect to the right-handed neutrinos, so that the effective
light neutrino mass matrix is $m_\nu =m_D
\frac{1}{M}\mu\frac{1}{M}m_D^T$. Assuming $m_D$ and $\mu$ to be
proportional to the identity matrix and $M\sim M_{TBM}$, it is
straightforward to show that we can obtain the mass sum-rule of type
{\it (D)}.


A novel seesaw mechanism arises from left-right
symmetry~\cite{Akhmedov:1995vm} or the full
SO(10)~\cite{malinsky:2005bi} in the presence of gauge singlet
fermions, and has been called the linear seesaw. In such scheme the
effective light neutrino mass matrix is given in terms of three
independent sub-matrices and scales linearly with respect to the usual
Dirac neutrino Yukawa couplings, hence the name. In order to have a
mass relation, we need two sub-matrices of numerical ``TBM-type'' like
in eq.(\ref{tbmtype}) and the third one similar to $M_{TBM}$ (see also
\cite{Hirsch:2008rp}), otherwise additional free parameters are
introduced, beyond our assumed two.  One can show that all four cases
can be realized, depending on which matrix has the form $M_{TBM}$.

\section{Lower bound for neutrinoless double-$\beta$ decay}

\begin{figure}[h!]
  \begin{center}
 \includegraphics[width=75mm,height=55mm]{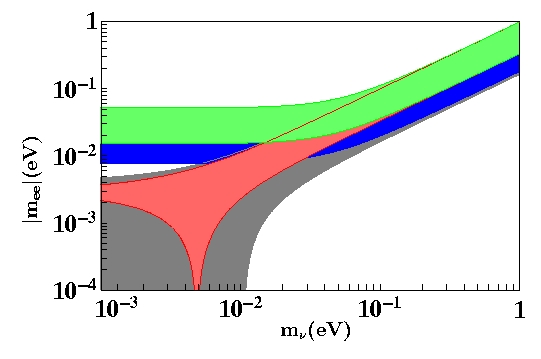}
 \end{center}
 \caption{Comparison on the allowed range of $\langle |m_{ee}|\rangle$
   as a function of the lightest neutrino mass. For the TBM mixing
   pattern (red and green bands for NH and IH respectively) and for
   the full allowed 3$\sigma$ C.L. ranges of oscillation parameters
   from~\cite{Schwetz:2011qt,Schwetz:2011zk} (gray and blue bands for
   NH and IH respectively). }
 \label{fig2}
\end{figure}

Let us first consider the amplitude for neutrinoless double-$\beta$ decay
within a flavor-generic model.
One can plot the effective neutrino mass parameter $|m_{ee}|$
determining the $0\nu\beta \beta$ decay amplitude, as a function of
the lightest neutrino mass. As is well-known, by varying the neutrino
oscillation parameters $\Delta m^2_{atm}$, $\Delta m^2_{sol}$,
$\theta_{12}$, $\theta_{13}$, $\theta_{23}$ in their allowed
ranges~\cite{Schwetz:2011qt,Schwetz:2011zk} one obtains two types of
relatively broad bands in the $(|m_{ee}|,m_{\text{light}}^\nu)$ plane
corresponding to normal and inverse hierarchy spectra, as shown in
Fig.~\ref{fig2}.

In this ``generic'' case there is a lower bound on the neutrinoless
double-$\beta$ decay effective mass parameter $|m_{ee}|$ only in the
case of inverse mass hierarchy: due to the possibility of destructive
interference among the light neutrinos from the effect of having
opposite CP signs or due to the effect of Majorana phases, no lower
bound can be established for the case of normal
hierarchy~\cite{wolfenstein:1981rk,schechter:1981hw,valle:1983yw}.

Let us now turn to the case where MSR relations like {\it (A),(B),(C)}
and {\it (D)} hold. As discussed above these can be obtained in
flavor models where the neutrino mass matrix only depends on two
independent free parameters, so that the resulting mixing angles are
fixed, like for example for the tri-bimaximal or bimaximal mixing
patterns.

For definiteness here we focus on the case where the rotation in the
neutrino sector is of tri-bimaximal form. Corrections from higher
dimensional operators and/or from the charged lepton sector can yield
$\theta_{13} \neq 0$, as suggested after the T2K~\cite{Abe:2011sj} and
Double-Chooz~\cite{Double-Chooz} first results~\cite{Schwetz:2011qt}.

Hence we retain the TBM approximation as a useful starting point to
obtain our MSR relations. However, when evaluating a lower bound on
the effective neutrino mass parameter $|m_{ee}|$ determining the
neutrinoless double-$\beta$ decay amplitude, we include explicitly the
effects of non-vanishing $\theta_{13}$. We do this by taking the
values at $3~\sigma$ determined in Ref.~\cite{Schwetz:2011qt}.  Such a
lower bound can be obtained from the following procedure.

\begin{figure}[h!]
  \begin{center}
 \includegraphics[scale=.3]{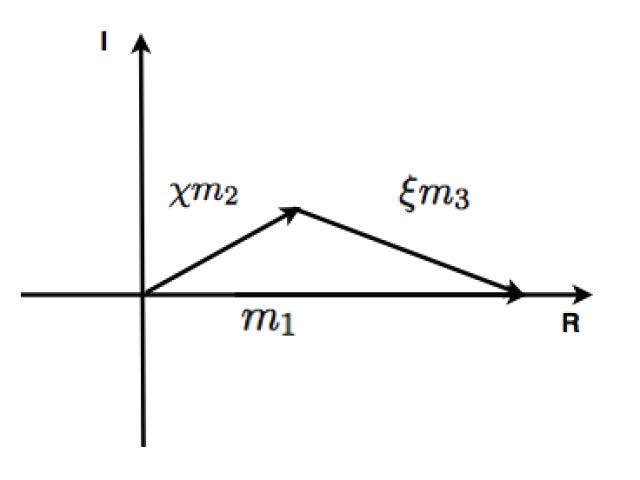}
 \end{center}
\caption{Neutrino mass sum-rule triangle in the complex plane.}
  \label{triangle}
\end{figure}

We first consider that the neutrino masses are complex parameters,
where the two Majorana phases are encoded in $m_2^\nu$ and
$m_3^\nu$,~i.e.
\begin{eqnarray}
&m_1^\nu&=m_1^0, \label{m1}\\
&m_2^\nu&=m_2^0 e^{i\alpha},\label{m2}\\
&m_3^\nu&=m_3^0 e^{i\beta}. \label{m3}
\end{eqnarray}
As shown in Fig.~\ref{triangle}, the neutrino mass sum-rule can then
be interpreted geometrically as a triangle in the complex plane, whose
area provides a measure of Majorana CP violation~\footnote{As the area
  shrinks to zero one obtains the CP-conserving limits corresponding
  to the four independent choices of CP
  sign~\cite{wolfenstein:1981rk,schechter:1981hw}.}.
Each of the above equations~(\ref{m1}),~(\ref{m2}) and (\ref{m3}) can
be split into two independent equations for the real and imaginary
parts.

For simplicity let us start from the idealized case where the neutrino
oscillation parameters $\Delta m^2_{atm}$, $\Delta m^2_{sol}$,
$\theta_{12}$, $\theta_{13}$, $\theta_{23}$ are perfectly
well-measured quantities. In this case one can extract the two
Majorana phases $\alpha$ and $\beta$ as functions of the base of the
triangle, which is determined by $m_1^0$ in case of normal hierarchy
(NH) or by $m_3^0$ in case of inverted hierarchy (IH), as well as the
parameters $\chi$ and $\xi$ labeling the particular model under
consideration.

These relations obtained can then be inserted into the general
expression of $|m_{ee}|$:
\begin{equation}
\label{0nubb}
|m_{ee}|  = \left|c_{12}^2c_{13}^2m_1+s_{12}^2c_{13}^2e^{i\alpha}m_2+s_{13}^2e^{i\beta}m_3\right| .
\end{equation}
For each ($\chi$, $\xi$) model this effective mass parameter depends on a
single parameter, namely the length of the triangle base, which
gives a measure of the absolute scale of neutrino mass.

For instance, for case {\it (A)} this procedure gives:
\begin{equation}
\xi \cos\alpha\, m_2^0 +\chi \cos\beta \,m_3^0 =m_1^0,\quad
\xi \sin\alpha \,m_2^0 +\chi \sin\beta \, m_3^0 =0 
\,,
\end{equation}
so that the Majorana CP phases are determined as:
\begin{eqnarray}
\cos\alpha&=&\frac{m_1^2-\chi ^2 \left(\Delta m^2_{atm}+m_1^2\right)+\xi^2 \left(\Delta m^2_{sol}+m_1^2\right)}
{2 m_1 \xi  \sqrt{ \Delta m^2_{sol}+m_1^2}} \nn \\ \label{esempio}\\
\cos\beta&=&\frac{m_1^2 + \chi ^2 \left(m_1^2+\Delta m^2_{atm}\right)-\xi^2 \left(m_1^2+\Delta m^2_{sol}\right)}
{2 m_1 \chi  \sqrt{ \Delta m^2_{atm}+m_1^2}}\,. \nn
\end{eqnarray}
The lower bound for the lightest neutrino mass can be obtained from
our MSR, using the triangle inequality in the complex plane as
suggested by Rodejohann and Barry in \cite{Barry:2010yk}, see
Fig.(\ref{triangle}) for a schematic view.

We must first select the biggest side of the triangle; calling them
$x_1$, $x_2$ and $x_3$, then the triangle inequality $|x_i|\leq
|x_j|+|x_k|$ must be fulfilled, where $|x_i| \equiv
\text{Max}(|x_1|,|x_2|,|x_3|)$ and $i\ne j\ne k$~\footnote{Notice that
  there are three inequalities of the type $|x_i|\leq |x_j|+|x_k|$
  obtained by permuting the three indices $i, \,j$ and $k$, but only
  one of these constrains the lightest neutrino mass.}.  In case {\it
  (A)} and assuming NH for the neutrino mass spectrum, we always have
$(\chi m_2^0,\xi m_3^0 ) > m_1^0$ and the largest side of the triangle
can be either $\chi m_2^0$ or $\xi m_3^0$; so we must consider
separately these two cases.  After rewriting two masses in terms of
the two squared mass differences, we can obtain a lower limit for the
lightest neutrino mass from the triangle inequality $|x_i|\leq
|x_j|+|x_k|$.  For the other cases we follow the same procedure.
The lower bound on the lightest neutrino mass obtained in this way is
then used to estimate the lower bound for $|m_{ee}|$ from the general
expression in eq.~(\ref{0nubb}). Notice that, although we focus here
on TBM schemes, some of the MSR considered in our analysis can also be
derived using bimaximal mixing as a starting point: for instance, in
Refs.~\cite{Altarelli:2009gn,Meloni:2011fx} a relation of type $(A)$
with $(\chi,\xi)=(1,2)$ has been derived and the phenomenological
consequences studied.

\subsection*{Classification}

\begin{figure}[h!]
  \begin{center}
 \includegraphics[width=70mm,height=55mm]{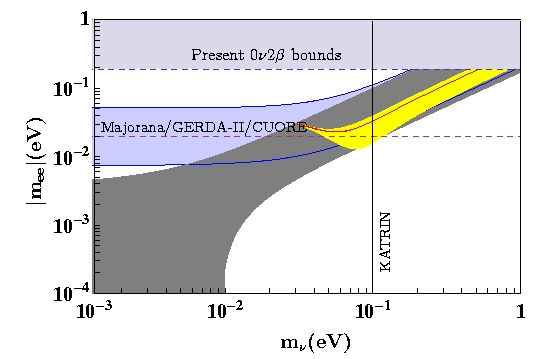}
 \includegraphics[width=70mm,height=55mm]{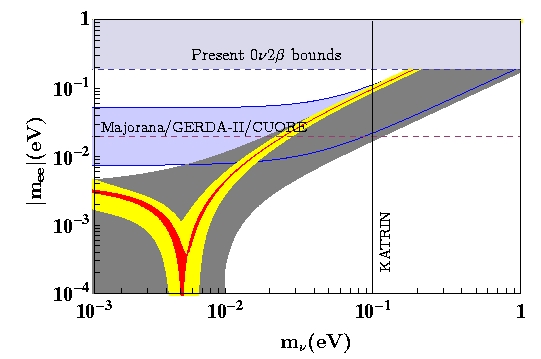}
 \end{center}
 \caption{$\langle |m_{ee}|\rangle$ as a function of the lightest
   neutrino mass corresponding to the mass sum-rule $2\sqrt{m_2} +
   \sqrt{m_3} = \sqrt{m_1}$ \cite{Hirsch:2008rp} (left) and
   $3\sqrt{m_2} + 3\sqrt{m_3} = \sqrt{m_1}$ (right).  The red bands
   correspond to the TBM mixing pattern, while the yellow bands
   correspond to the same MSR, but now varying the values of
   $\theta_{13}$, $\theta_{23}$ and $\theta_{12}$ to 3$\sigma$
   C.L. range.}
  \label{fig3}
\end{figure}

\begin{figure}[h!]
\centering
\includegraphics[width=70mm,height=55mm]{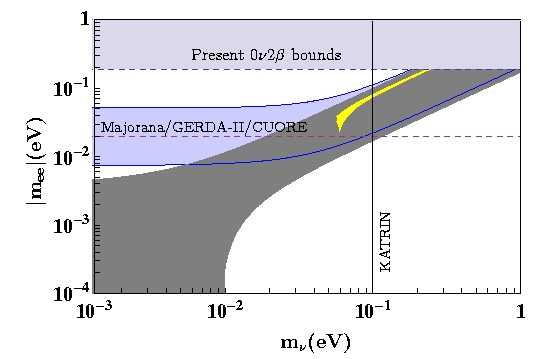}
\includegraphics[width=70mm,height=55mm]{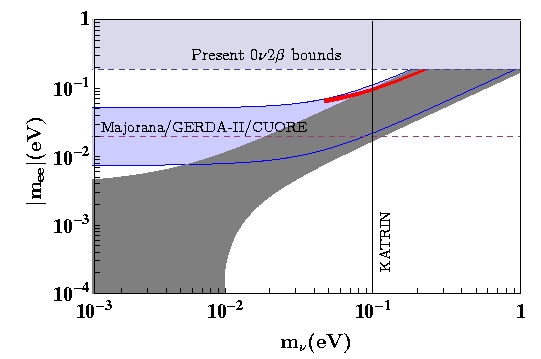}
\caption{Effective \znbb mass parameter $\langle |m_{ee}|\rangle$ as a
  function of the lightest neutrino mass for MSR type-A schemes. On
  the left panel the yellow band corresponds to the model which
  predicts the sum mass rule $3m_2 + 3m_3 = m_1$ for the case of
  NH. On the right the red band corresponds to the same sum rule in
  the case of IH. }
\label{fig4}
\end{figure}

\begin{figure}[h!]
\includegraphics[width=70mm,height=55mm]{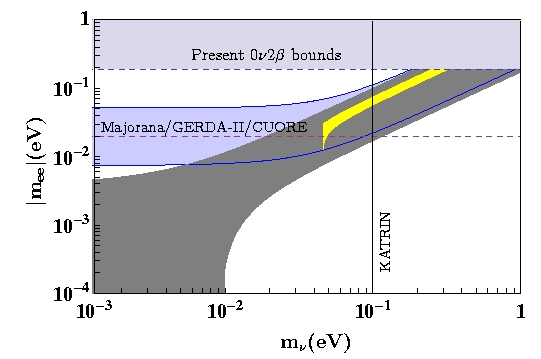}
\includegraphics[width=70mm,height=55mm]{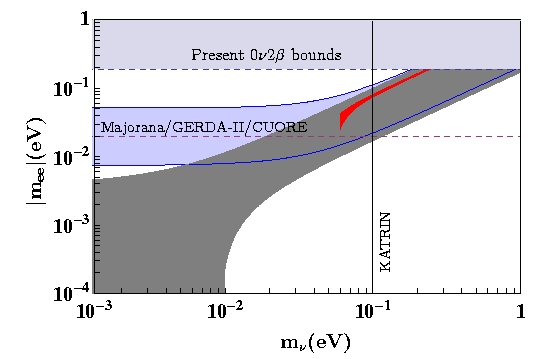}
\caption{$\langle |m_{ee}|\rangle$ as a function of the lightest
  mass. The figures correspond to the case (B). On the left panel the
  yellow band corresponds to the model which predicts the sum mass
  rule $1/m_2 + 3/m_3 = 1/m_1$ for the NH case. On the right, the red
  band corresponds to the prediction of the sum rule $3/m_2 + 3/m_3 =
  1/m_1$ in the case of IH. }
\label{fig5}
\end{figure}

\begin{figure}[h!]
\includegraphics[width=70mm,height=55mm]{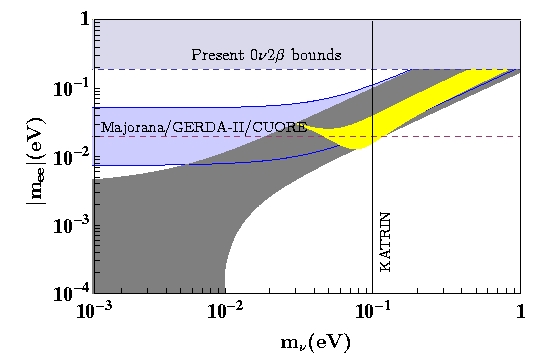}
\includegraphics[width=70mm,height=55mm]{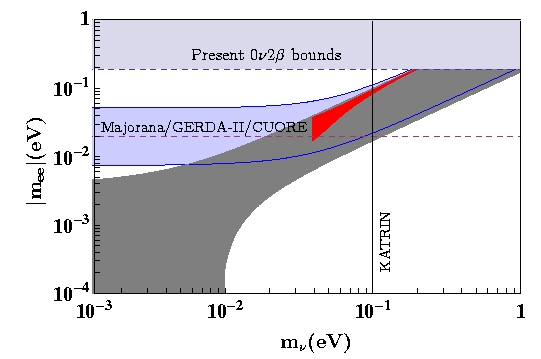}
\caption{$\langle |m_{ee}|\rangle$ as a function of the lightest
  neutrino mass. The figures correspond to the case (C). On the left
  the yellow band corresponds to the model which predicts the sum mass
  rule $3\sqrt{m_2} + 3\sqrt{m_3} = \sqrt{m_1}$ in case of NH. On the
  right, the red band corresponds to the prediction of the sum mass
  rule $2\sqrt{m_2} + 3\sqrt{m_3} = \sqrt{m_1}$ in the case of IH. }
\label{fig6}
\end{figure}

\begin{figure}[h!]
\includegraphics[width=70mm,height=55mm]{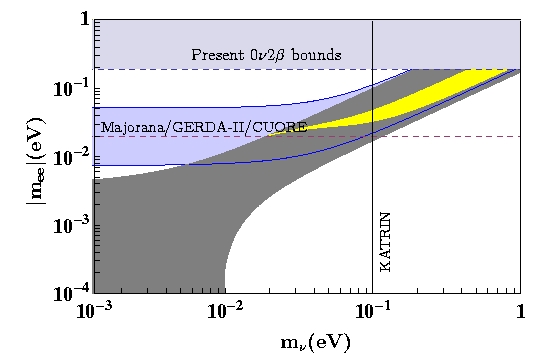}
\includegraphics[width=70mm,height=55mm]{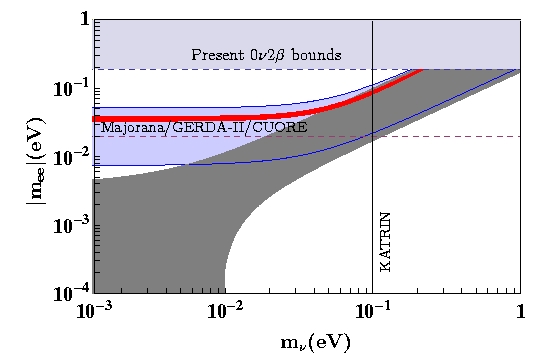}
\caption{$\langle |m_{ee}|\rangle$ as a function of the lightest
  neutrino mass, case (D). On the left the yellow band corresponds to
  the MSR scheme which predicts $3/\sqrt{m_2} + 3/\sqrt{m_3} =
  1/\sqrt{m_1}$ for NH. On the right panel, the red band corresponds
  to the prediction of the mass sum rule $1/\sqrt{m_2} + 2/\sqrt{m_3}
  = 1/\sqrt{m_1}$ in the case of IH. }
\label{fig7}
\end{figure}

\begin{table}[h!]
\begin{center}
\begin{tabular}{|c||c|c|c||c|c|c||c|c|c||c|c|}
\hline
$\chi,\xi$ &A--NH& A--IH&Ref.& B--NH& B -- IH&Ref.&  C-- NH& C --IH&Ref.&  D --NH& D-- IH\\
\hline
\hline
1,1&0.010 &0.044&\cite{Ma:2005sha,Altarelli:2005yp,Altarelli:2005yx,Altarelli:2006kg,Ma:2006vq,Bazzocchi:2007na,Bazzocchi:2007au,Honda:2008rs,
Brahmachari:2008fn,Lin:2008aj,Chen:2009um,Ma:2009wi,Fukuyama:2010mz,Bazzocchi:2008ej,Chen:2007afa,Ding:2008rj,Chen:2009gy}&0.008&0.036&
\cite{Barry:2010zk,Ding:2010pc,Bazzocchi:2009da}&0.006 & 0.029& -&.005&.008\\
\hline
1,2&$\star$&0.046&-&0.008&0.027&-&$\star$ &0.014& -&.004&.026\\
\hline
1,3&$\star$ &0.011&-&0.030&0.005&-&$\star$ &0.014& -&.018&.025\\
\hline
2,1&0.006 &$\star$ &\cite{Ma:2005sha,Altarelli:2005yp,Altarelli:2005yx,Altarelli:2006kg,Ma:2006vq,Bazzocchi:2007na,Bazzocchi:2007au,Honda:2008rs},&0.006&0.007&
\cite{Altarelli:2005yp,Altarelli:2005yx,Chen:2009um},&0.000 & $\star$& \cite{Hirsch:2008rp}&$\star$&.007\\
&&&\cite{Brahmachari:2008fn, Bazzocchi:2009pv}&&&
\cite{Chen:2009gy,Burrows:2010wz,Babu:2005se,He:2006dk,Morisi:2007ft,Altarelli:2008bg,Adhikary:2008au,Csaki:2008qq,Altarelli:2009kr,Lin:2009bw,
Hagedorn:2009jy,Burrows:2009pi,Berger:2009tt,Ding:2009gh,Mitra:2009jj,delAguila:2010vg}&&&&&\\
\hline
2,2&0.019&0.026&-&0.023&0.008&-&0.017&$\star$&-&.003&.015\\
\hline
2,3&$\star$&0.046&-&0.007&0.008&-&$\star$ & 0.031&-&.005&.026\\
\hline
3,1&0.004&$\star$&-&0.004&0.008&-&$\star$ & $\star$&-&$\star$&$\star$\\
\hline
3,2&0.011&$\star$&-&0.004&0.021&-&0.000 & $\star$&-&$\star$&.007\\
\hline
3,3&0.023 &0.061&-&0.029&0.031&-&0.011 & 0.019&-&.018&.016\\
\hline
\end{tabular}\caption{Minimal values for the effective  $0\nu\beta \beta$
  decay mass parameter $|m_{ee}|$, in eV, see text for details.}\label{tab1}
\end{center}
\end{table}
The results obtained from the procedure discussed above are summarized
in Tab.\ref{tab1}, where we report the lower limits of $|m_{ee}|$
corresponding to different integer choices of $(\chi,\xi)$ between 1
and 3 and for each of the four MSR considered in
eqs.~(\ref{i})-(\ref{iiii}), for both normal and inverted hierarchies
(compare also with Figs.~\ref{fig4}-\ref{fig7}).
The cases already discussed in the literature are indicated by giving
the corresponding reference. The entries denoted with the symbol
($\star$) represent situations that do not satisfy the inequality for
any value of the lightest neutrino mass $m_{light}$. Cases marked by a
(-) correspond to models which, as far as we can tell, have not been
considered.
Some comments are in order. First let us consider the effect of a
possible non-zero effect of $\theta_{13}$ as indicated by recent
experiments~\cite{Abe:2011sj,Double-Chooz} as well as global neutrino
oscillation fits~\cite{Schwetz:2011qt,Schwetz:2011zk,Fogli:2011qn}.
In Fig \ref{fig3} we show the prediction for $|m_{ee}|$ as function of
$m_{light}$ obtained from the MSR $3\sqrt{m_2} + 3\sqrt{m_3} =
\sqrt{m_1}$ (right panel) and $2\sqrt{m_2} + \sqrt{m_3} = \sqrt{m_1}$
(left panel). For the red bands we assumed the TBM values of the
oscillation parameters (implying $\theta_{13}=0$) while in the yellow
bands corresponds to the same MSR, but now varying the values of
$\theta_{13}$, $\theta_{23}$ and $\theta_{12}$ within their 3$\sigma$
C.L. interval.
By looking at the left panel in Fig.~\ref{fig3} one sees that, indeed,
the \znbb lower bound is sensitive to the value of $\theta_{13}$.

One also finds that, as expected on general grounds, all inverse
hierarchy schemes corresponding to various choices of $(\chi,\xi)$
within sum-rules A-D have a lower bound for the parameter
$|m_{ee}|$. However, the numerical value obtained depends on the MSR
scheme, signaling that not all values within the corresponding band
in Fig.~\ref{fig2} are covered.

On the other hand, even though normal hierarchy models do not lead to
a lower bound on the \znbb amplitude due to the possibility of
destructive interference amongst the light neutrinos, one finds that
the possibility of full cancellation is precluded for all schemes in
the table, except for the (2,1) case considered in
Ref.~\cite{Hirsch:2008rp} and the (3,2) scheme, both of which
correspond to MSR of type (C). All other NH MSR schemes considered
here imply a minimum value for the $0\nu\beta \beta$ decay
amplitude~\footnote{Of course some of the bounds are
  phenomenologically less interesting since they fall outside
  realistic sensitivities of coming experiments.}. One finds that the
most favorable cases are given by:
\begin{center}
\begin{tabular}{l}
$(\chi,\xi)=(3,3)$ for the case {\it (A)} NH,\\
$(\chi,\xi)=(3,3)$ for the case {\it (A)} IH,\\
$(\chi,\xi)=(1,3)$ for the case {\it (B)} NH,\\
$(\chi,\xi)=(3,3)$ for the case {\it (B)} IH,\\
$(\chi,\xi)=(3,3)$ for the case {\it (C)} NH,\\
$(\chi,\xi)=(2,3)$ for the case {\it (C)} IH,\\
$(\chi,\xi)=(3,3)$ for the case {\it (D)} NH,\\
$(\chi,\xi)=(1,2)$ for the case {\it (D)} IH.
\end{tabular}
\end{center}
In particular, the maximal value we have found for the lower bound on
$|m_{ee}|$ is $|m_{ee}|=0.061\,$eV, obtained in correspondence with
the set of values $(\chi,\xi)=(3,3)$ for the case {\it (A)} in IH. 
Such a value for $|m_{ee}|$ lies within the sensitivity of upcoming
experiments; hence it would be interesting, from the model building
point of view, to find from first principles a flavor-symmetry-based
model predicting such a mass relation; we will return to this problem
elsewhere.

The same phenomenologically interesting cases are now studied more in
detail, showing the behavior of $|m_{ee}|$ as function of the
lightest neutrino mass in Figs.~\ref{fig4},\ref{fig5},\ref{fig6} and
\ref{fig7}.  In all plots, the two bands are the most generic ones
compatible with both normal and inverted hierarchies, derived
considering the 3$\sigma$ allowed ranges on the neutrino oscillation
parameters as obtained in Ref.\cite{Schwetz:2011qt} and consistent
with latest T2K and Double-Chooz experiments, see Fig. \ref{fig2}.

In Fig.~\ref{fig4} we give the allowed $\langle |m_{ee}|\rangle$
values as a function of the lightest neutrino mass. The figures
correspond to case (A). In the left panel the yellow band corresponds
to the model which predicts the mass sum rule $3m_2 + 3m_3 = m_1$ in
case of NH. On the right the red band corresponds to the same sum rule
in the case of IH. Other MSR \znbb amplitude lower bounds are
illustrated in subsequent figures.


\section{Conclusions}
\label{c}

In this paper we have analyzed the implications for the lower bound on
the effective \znbb neutrino mass parameter $|m_{ee}|$ arising from
possible mass sum-rules obtained in the context of flavor models.
Mass sum rules are classified in four different categories, some have
already been considered in the literature.  For each case, we have
first extracted the allowed numerical values of $|m_{ee}|$, for both
mass orderings of the neutrino mass eigenstates and we have then given
the behavior of $|m_{ee}|$ as a function of the lightest neutrino
mass. Although our MSR schemes were obtained within the TBM anzatz, we
have computed the possible values of $|m_{ee}|$ considering all the
neutrino parameters (including a non-vanishing $\theta_{13}$) within
their 3$\sigma$ allowed ranges.  In most MSR schemes one finds a lower
bound for the \znbb amplitude, even for NH spectra.
We find that the most favorable case (large lower bound) corresponds
to a sum-rule of type $(A)$ obtained in correspondence of the set of
values $(\chi,\xi)=(3,3)$, $|m_{ee}|=0.061\,$eV.  Such a mass relation
has not been considered so far, and the searching of a flavor model
able to predict it at leading order is now in progress.

\section{Acknowledgments}

This work was supported by the Spanish MICINN under grants
FPA2008-00319/FPA, FPA2011-22975 and MULTIDARK CSD2009-00064
(Consolider-Ingenio 2010 Programme), by Prometeo/2009/091 (Generalitat
Valenciana), by the EU ITN UNILHC
PITN-GA-2009-237920. S. M. is supported by a Juan de la Cierva
contract. E. P. is supported by CONACyT (Mexico).  D.M. acknowledges
MIUR (Italy), for financial support under the contract PRIN08.


\end{document}